\documentclass[letterpaper,12pt,final]{iopart}
\pdfoutput=1
\usepackage{braket}
\usepackage{graphicx}
\usepackage{subfigure}
\usepackage{wrapfig}

\usepackage[usenames, dvipsnames]{color}
\usepackage[normalem]{ulem}

\begin{document}
\title[Study of the polarisation of annihilation photons using Compton scattering]{An undergraduate laboratory study of the polarisation of annihilation photons using Compton scattering}

\author{P.~Knights$^{1}$, F.~Ryburn$^{2}$\footnote[1]{Ogden Trust Summer Intern at the University of Birmingham.}, G.~Tungate$^{1}$, K.~Nikolopoulos$^{1}$}

\address{$^{1}$ School of Physics and Astronomy, University of Birmingham, B15 2TT, United Kingdom}
\address{$^{2}$ School of Physics and Astronomy, University of Oxford, OX1 3RH, United Kingdom}
\ead{k.nikolopoulos@bham.ac.uk}

\begin{abstract}
An experiment for the advanced undergraduate laboratory which allows
students to study the effect of photon polarisation in Compton
scattering and to explore quantum entanglement is described. The
quantum entangled photons are produced through electron-positron
annihilation in the $S$-state, and their polarisations are analysed
using the Compton scattering cross-section dependence on the photon
polarisation. The experiment was equipped with off-the-shelf detectors
and electronic units.  Finite geometry effects are discussed and
investigated with the use of a Geant4-based simulation.
\end{abstract}

\submitto{\EJP}

\section{Introduction}
Entanglement is an intriguing quantum mechanical phenomenon and the
subject of the famous Einstein-Podolsky-Rosen (EPR)
Paradox~\cite{Einstein1935}. It was shown that a correlation exists
between the quantum states of particles which had previously
interacted even when they are no longer interacting. This leads to the
classically paradoxical situation of apparent communication between
particles separated by space-like intervals. This experimentally
verified result of quantum
mechanics~\cite{PhysRevLett.28.938,PhysRevLett.47.460,PhysRevLett.49.1804}
has profound implications in many fields including quantum computing
and encryption.

An experiment is presented in which the effect of photon entanglement
is explored. A pair of entangled photons, polarised at right angles to
each other, are produced from positronium annihilation. Positronium,
the bound state of an electron and a positron, is dominantly formed in
a singlet $S$-state, where the electron and positron spins are
anti-parallel~\cite{Wheeler1946}. This state has a mean lifetime of
approximately 0.125~ns and decays preferentially via annihilation to
produce two photons, each with an energy of 511~keV traveling in
opposite directions. For angular momentum conservation the two photons
must be produced in a polarisation state
\begin{equation}
    \phi = \frac{\ket{xy}-\ket{yx}}{\sqrt{2}}
    \label{eq:wavefunction}
\end{equation}
where $\ket{xy}$ ($\ket{yx}$) denotes the state where the first photon
is polarised in the x-plane (y-plane) and the second in the y-plane
(x-plane)~\cite{Dirac1930,Snyder1948}.
Subsequently, the two photons travel to two ``scatterers'' where they
undergo Compton scattering~\cite{Compton:1923zz}. Given the entangled
nature of the photons the interaction of the first determines the
polarisation of the second. This connection, combined with the
sensitivity of Compton scattering to photon polarisation, results in
an angular correlation of the scattered photons. Having fixed the
scattering plane of one of the photons this correlation is quantified
by an asymmetry in the number of times the other photon scatters into
planes parallel and perpendicular to the fixed plane. This angular
correlation persists regardless of the distance between the two
``scatterers''.  Similar experiments involving Compton scattered
annihilation photons have been proposed~\cite{Wheeler1946},
conducted~\cite{Wu:1950zz,Kasday1975,Wilson1976}, and calculations
have been carried out~\cite{Pryce1947,Snyder1948}. Recently, a similar
experiment was proposed based on semiconductor detectors for lecture
demonstration of the photon entanglement~\cite{Hetfleis2017}.
 
Although the EPR paradox was a topic of extensive experimental study
in the mid--twentieth century, it remains mostly absent from the
undergraduate laboratory. Experiments to study photon entanglement
have been proposed~\cite{Dehlinger2002,Mitchell2002} using photons
near the visible range of the electromagnetic spectrum. Nevertheless,
the sensitivity of Compton scattering on the photon polarisation
provides the means to extend such studies to $\gamma$-rays. This
experiment complements a series of experiments on Compton scattering
using $\gamma$-rays that has been an important part of the
undergraduate laboratory for many decades. Various forms of such
experiments have been proposed and implemented, from the measurement
of absorption coefficients to probe the characteristics of the
scattered photons~\cite{Bartlett1964a}, to experiments involving
precision spectroscopy and timing
components~\cite{Bartlett1964d,French1965,Stamatelatos1972} and
experiments investigating polarisation effects~\cite{Knights:2017akf}.

\section{Experimental Arrangement}
\begin{figure}[h!]
\centering
\includegraphics[width=0.4\textwidth] {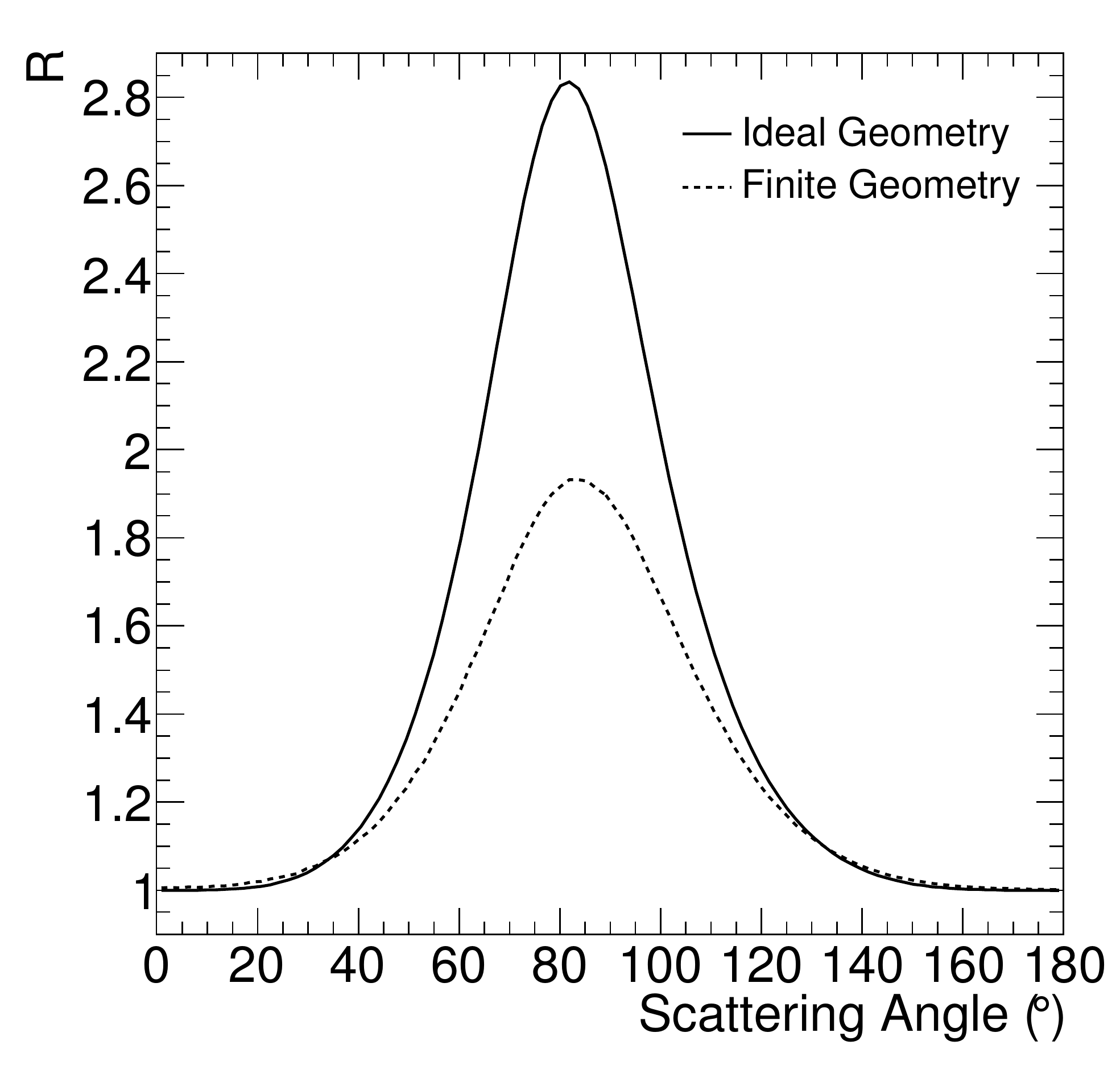}
\caption{The asymmetry ratio, R, as a function of scattering angle, $\theta$. The maximum of 2.84 is found at 82$^\circ$. The effect of finite geometry is also demonstrated for the case where the detectors subtend angles of $\pm 20^\circ$ and $\pm 25^\circ$ around the nominal $\theta$ and $\phi$ angles, respectively. \label{fig:asymmetry}}
\end{figure}

Consider two scatterers placed symmetrically about a positron
source. The photons released in annihilation will travel approximately
back-to-back. Upon reaching a scatterer, the first photon to interact
though Compton scattering fixes the polarisation plane. At that point
the wavefunction collapses, ensuring that the polarisation of the
second photon is perpendicular to that of the first. Due to the
polarisation dependence of Compton scattering this causes the other
photon to have a greater probability of scattering into a plane
perpendicular to that in which the first photon was scattered. The
possibility of photon interference is eliminated by setting the
distance between the two scatterers to be greater than the coherence
length, which for photons produced from $S$-state positronium decay is
approximately 4~cm.  The ratio of the number of photons scattered to
the perpendicular plane, $N_{\phi=\frac{\pi}{2}}$, to the number of
photons scattered to the parallel plane, $N_{\phi=0}$, is given
by~\cite{Snyder1948,Pryce1947}:
\begin{equation}
\label{eq:assymetry}
 {\rm R} = \frac{{\rm N}_{\phi=\frac{\pi}{2}}}{{\rm N}_{\phi=0}} = 1+\frac{2\sin^4\theta}{\gamma^2-2\gamma\sin^2\theta},
\end{equation}
where,
$\gamma=2-\cos\theta+\left(2-\cos\theta\right)^{-1}$,
with $\theta$ the scattering angle and $\phi$ the azimuthal angle defined between the planes of scattering. This is presented in Figure~\ref{fig:asymmetry} as a function of $\theta$, showing that for a scattering angle of 90$^\circ$ the expected ratio is 2.60. In an actual experiment the detectors will subtend finite angles which will lead to a smearing of the asymmetry. This is shown by the dotted line in Figure~\ref{fig:asymmetry} for the case where the detectors cover $\pm$20$^\circ$ and $\pm$25$^\circ$ around the nominal $\theta$ and $\phi$ angles, respectively. 

\begin{figure}[h!]
\centering
\subfigure[\label{fig:setupSketch}]{\includegraphics[width=0.45\textwidth]{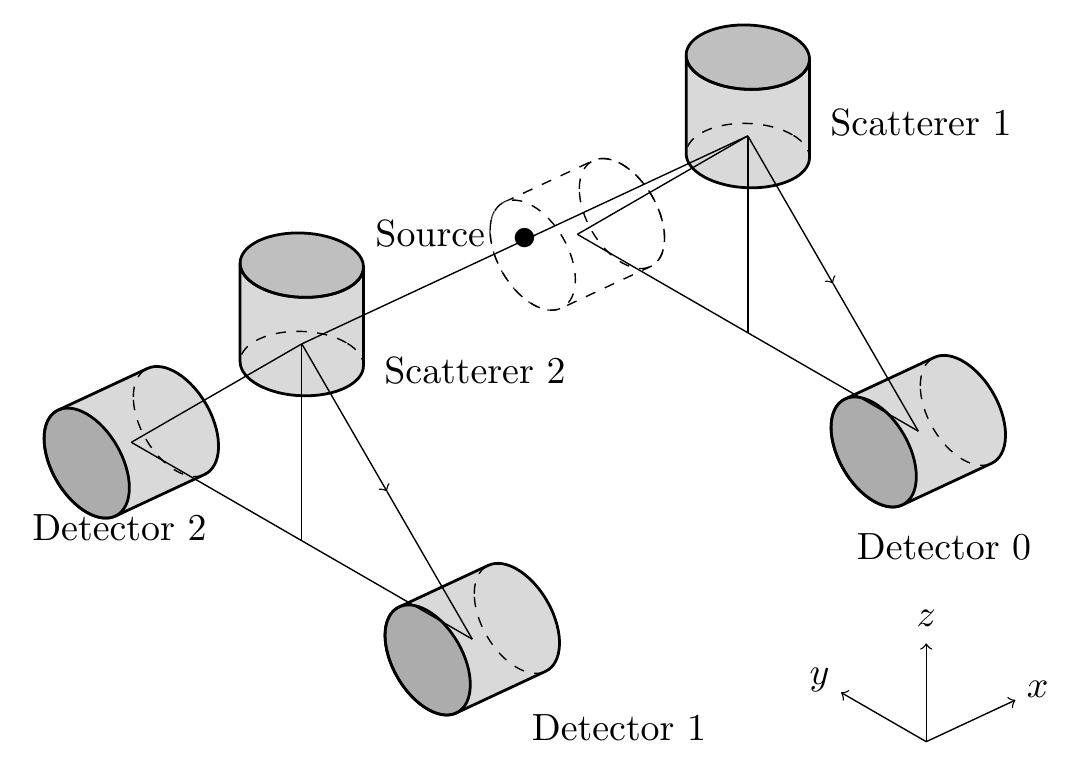}}
\subfigure[\label{fig:setup}]{\includegraphics[width=0.45\textwidth]{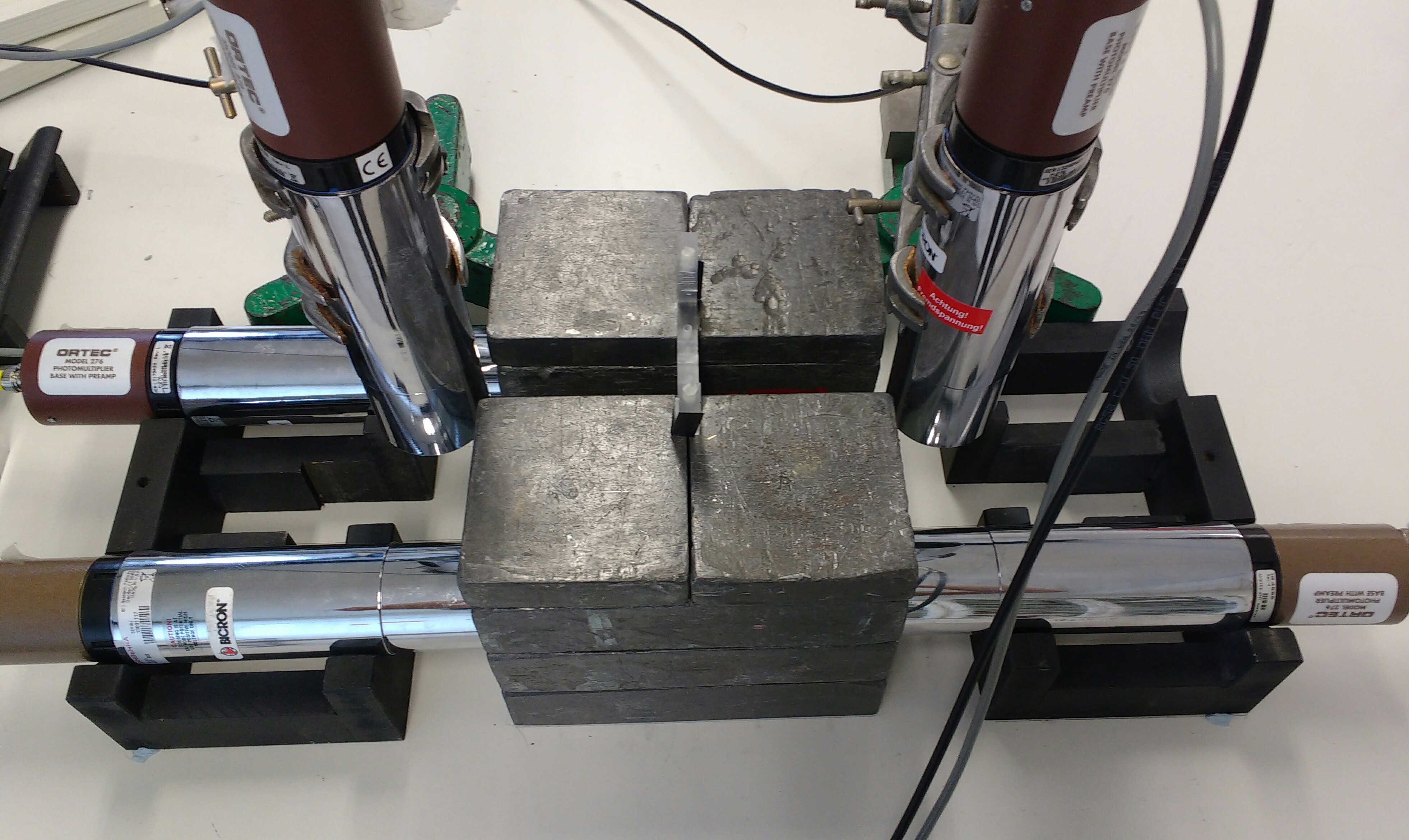}}
\caption{
    \subref{fig:setupSketch} Schematic diagram of the experimental set up showing the position of the source relative to the scatterers and detectors. 
    \subref{fig:setup} Photograph of the experimental set up showing the scatterers, detectors and the lead shielding.  
\label{fig:experiment}
}
\end{figure}

The experiment consists of five cylindrical NaI(Tl) scintillation
detectors, each 2'' in diameter and length. Annihilation photons are
obtained from a Germanium-68 source, which decays via electron capture
to Gallium-68. This subsequently decays to Zinc-68 emitting a
positron. The activity of the Germanium-68 source was approximately
4.4~kBq. Two cylindrical NaI(Tl) scintillators, Scatterer-1 and
Scatterer-2, are placed symmetrically about the source with their
faces at a distance of 16.2~cm.
The centres of the scatterers were arranged to be 1.1 cm above the
source along the $z$-axis to minimise the path a scattered photon
travels through the crystal before it emerges towards a detector. The
experimental layout is shown schematically in
Figure~\ref{fig:setupSketch}.

An annihilation photon emerging from the source impinges on
Scatterer-1 and is Compton scattered through 90$^\circ$ depositing
255.5~keV. The scattered photon, with energy 255.5~keV, then travels
11.4~cm to Detector-0 where it is absorbed and detected. The second
annihilation photon travels in the opposite direction and Compton
scatters in Scatterer-2 by 90$^\circ$. The photon may then impinge on
Detector-1 or Detector-2, where scattering to Detector-1 (Detector-2)
is in a plane parralel (perpendicular) to the scattering of the first
photon. The laboratory realisation of the experiment is shown in
Figure~\ref{fig:setup}, including the lead blocks used to shield the
detectors from photons arriving directly from the source.

From a Geant4~\cite{Agostinelli:2002hh} simulation of the experiment
it was obtained that 2.8\% of photons produced by an isotropic
positron source impinge on Scatterer-2. Of these, approximately 52\%
were fully absorbed, 43\% Compton scattered before leaving the
detector, while the remaining 5\% did not interact with the
detector. Approximately 0.013\% of photons from the source arrive in
Detector-1 having undergone Compton scattering in Scatterer-2.

\section{Detector Calibration and Characterisation}
Calibration of the detectors was performed using three
radio-isotopes~\cite{sonzogni2007nndc}: Germanium-68, Silver-108m and
Americium-241. Silver-108m decays by electron capture to
Palladium-108, emitting a number of photons, one being
434.0~keV. Americium-241 disintegrates by emmission of an alpha
particle to Neptunium-237, which releases a 59.5~keV photon. These
data were also used to estimate the energy resolution of Detector-1
and Detector-2, which was found to be approximately 7.0\% full-width
at half-maximum for 511~keV.  The timing of the detectors was
synchronised using annihilation photons.

The Germanium-68 source and a coincidence technique were used to
eliminate apparent asymmetries due to differences in efficiency
between Detector-1 and Detector-2. The detectors were placed
symmetrically about the source and detection of a signal in Detector-2
was gated by a signal compatible with a 511~keV photon in Detector-1
and vice-versa. The efficiencies of the two detectors were found to be
$\left(24.8\pm 0.5\right)\%$ and $\left(27.5\pm 0.6\right)\%$
respectively for 511~keV photons.

\section{Data Acquisition}
\label{sec:dataAcquisition}
\begin{figure}[h!]
\centering
\includegraphics[width=0.45\textwidth] {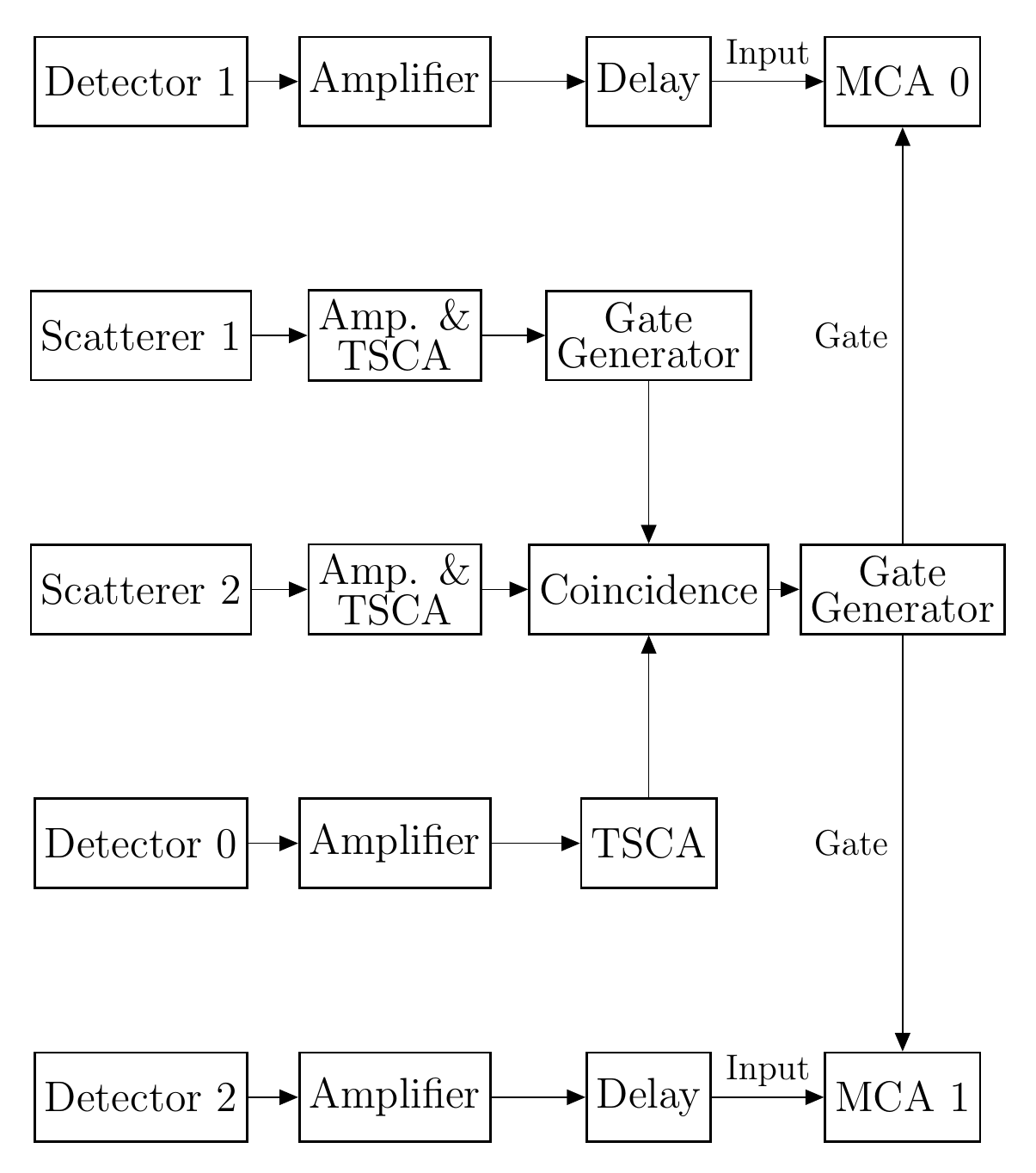}
\caption{Block diagram of the electronic components. \label{fig:setupElectronics}}
\end{figure}

The arrangement of electronic components for data acquisition is
summarised in Figure~\ref{fig:setupElectronics}.  The signals from
Detector-0, Scatterer-1, and Scatterer-2 were fed to three timing
single-channel analysers (TSCA). Energy windows for Detector-0,
Scatterer-1 and Scatterer-2 were set to accept signals corresponding
to energies between approximately 200~keV and 310~keV, using a method
similar to that employed in Ref.~\cite{Knights:2017akf}. Subsequently,
the TSCA signals were fed to a coincidence unit, with a resolving time
of approximately 2~${\rm \mu s}$, while the gate signal for Detector-1 and
Detector-2 had a width of approximately 3~${\rm \mu s}$.

Pulses from Detector-1 and Detector-2 were amplified, appropriately
delayed and each fed into a multi-channel analyser. The logic pulse
from the coincidence unit was used to gate the multi-channel
analysers.

The effect of the coincidence requirements on background was
examined. When requiring three-fold coincidence of Detector-0,
Scatterer-1 and Scatterer-2 a rate of approximately 8~${\rm s}^{-1}$
was observed in Detector-1 and Detector-2. Removing Scatterer-1 from
the coincidence does not significantly affect the rate, while removing
Scatterer-2 causes the rate to increase by a factor of two. Removing
both scatterers from coincidence leads to a further increase in the
rate by approximately an order of magnitude. To estimate the
background due to random coincidences further data were collected
where the coincidence signal was delayed by 20~${\rm \mu s}$.

\section{Results}
\begin{figure}[h!]
\centering
\subfigure[\label{fig:InitialSetup_Results1}]{\includegraphics[width=0.45\textwidth]{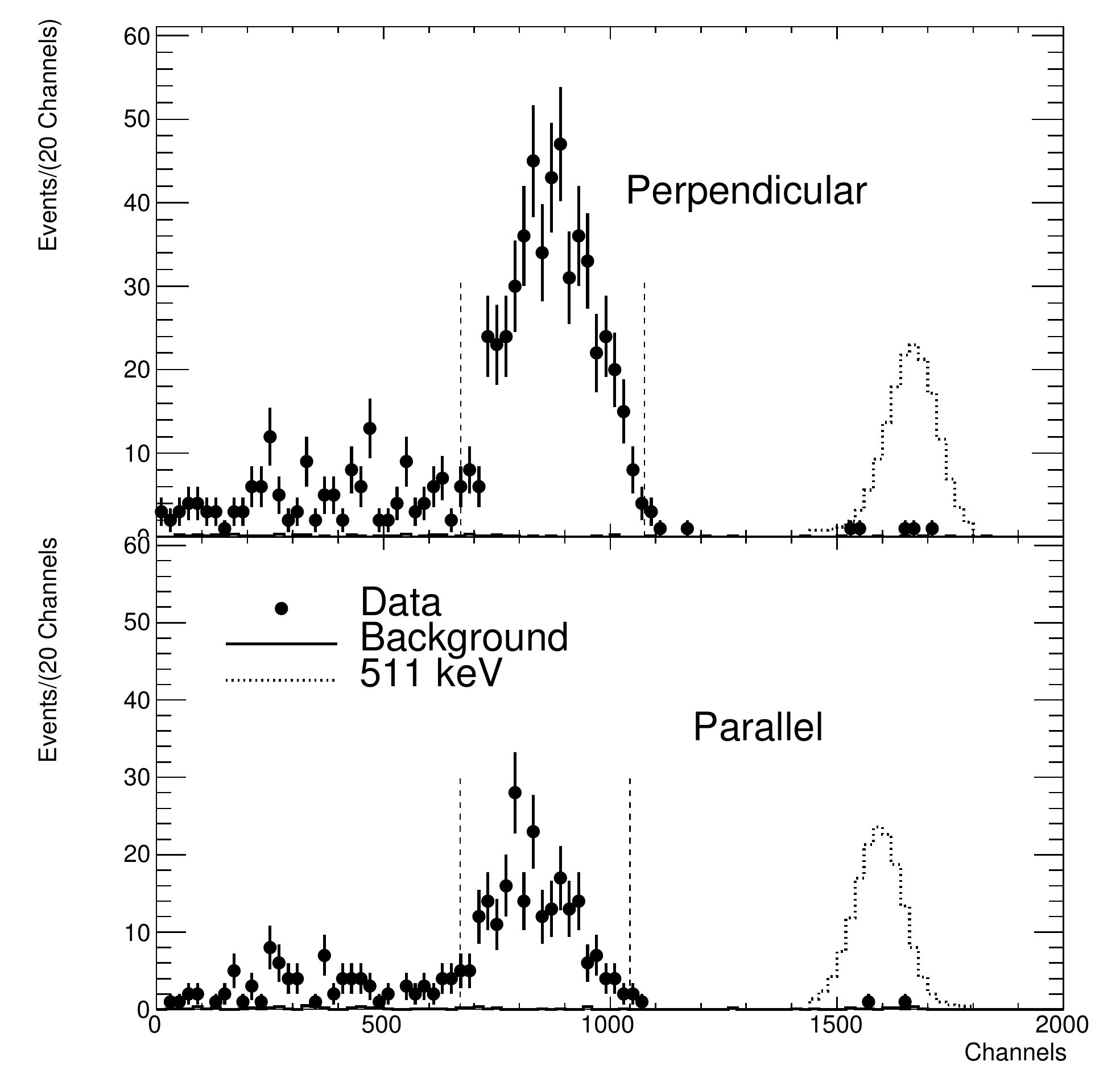}}
\subfigure[\label{fig:InitialSetup_Results2}]{\includegraphics[width=0.45\textwidth]{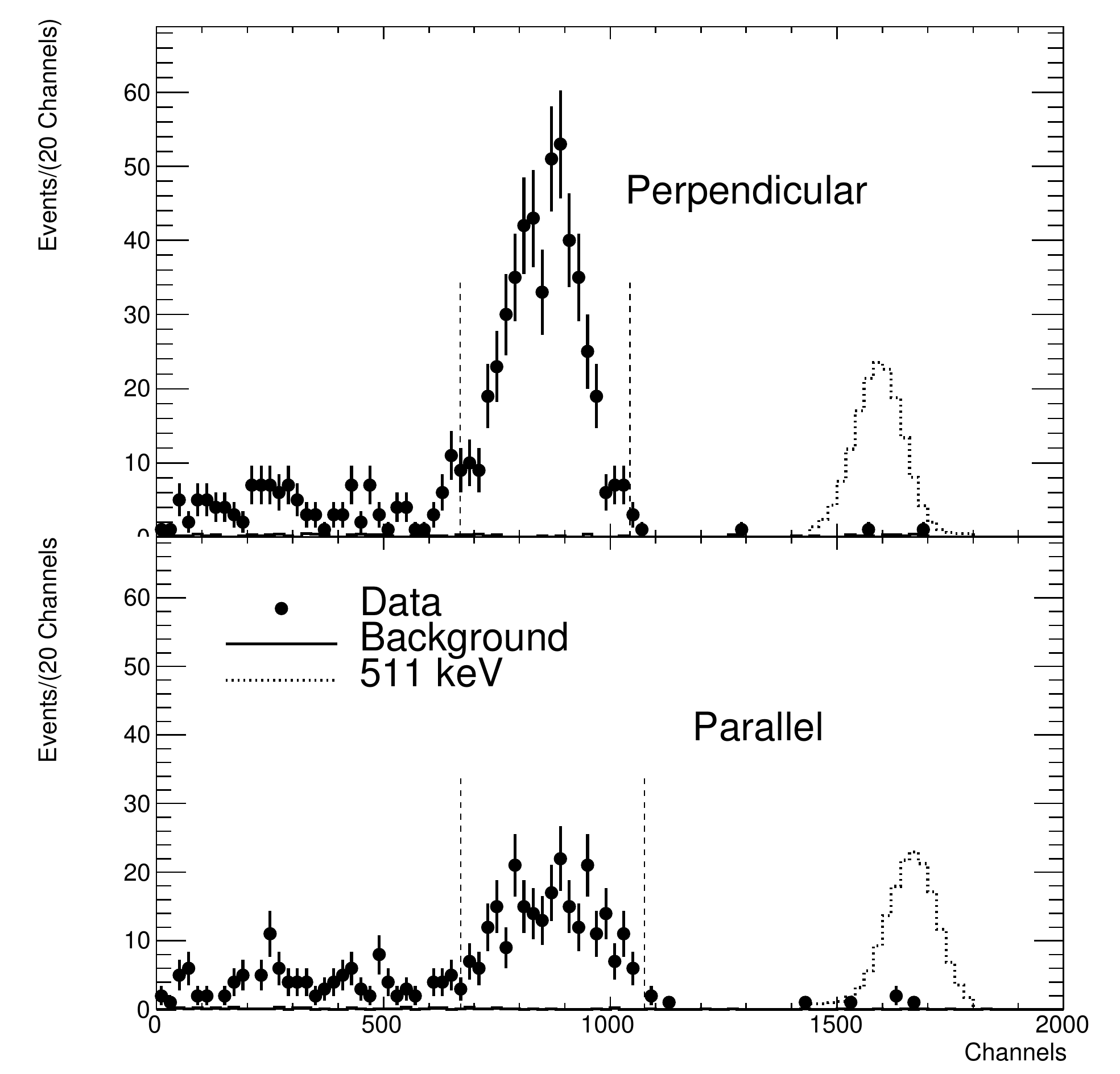}}
\caption{Results showing the asymmetry observed in the orthogonal planes for \subref{fig:InitialSetup_Results1} Detector-0 located as shown in Figure~\ref{fig:experiment}, and \subref{fig:InitialSetup_Results2} for Detector-0 moved to the location marked with a dashed cylinder. Perpendicular and Parallel refer to planes of scattering relative to scattering into Detector-0.\label{fig:results}
}
\end{figure}

The distribution of energies for the two detectors following data
collection for 20 hours is shown in
Figure~\ref{fig:InitialSetup_Results1} and an asymmetry is readily
observed. The peak at approximately ADC channel 850, corresponding to
an energy of 255~keV, is the full energy peak from the absorption of
the scattered photons. Additionally, X-rays from the excitation of
lead atoms of the shielding can be seen at approximately ADC channel
250, corresponding to 80~keV. The position of the 511~keV line from a
calibration run is also shown to set the energy scale.

To obtain the numerical results the peaks were integrated in the region demarcated by the dashed vertical lines, corresponding to an energy range of 201~keV to 327~keV. The small background due to accidental coincidences was then subtracted from this and the measured
ratio of perpendicular to parallel counts was taken giving a result of
$R_1=2.35\pm0.19$. The measurement was repeated by moving Detector-0, as shown with the dashed cylinder in Figure~\ref{fig:setupSketch}, which effectively interchanged the roles of Detector-1 and Detector-2 as parallel and perpendicular. The distribution obtained is shown in Figure~\ref{fig:InitialSetup_Results2} and results in $R_2=1.99\pm0.15$.

\begin{figure}[h!]
\centering
\includegraphics[width=0.60\textwidth]{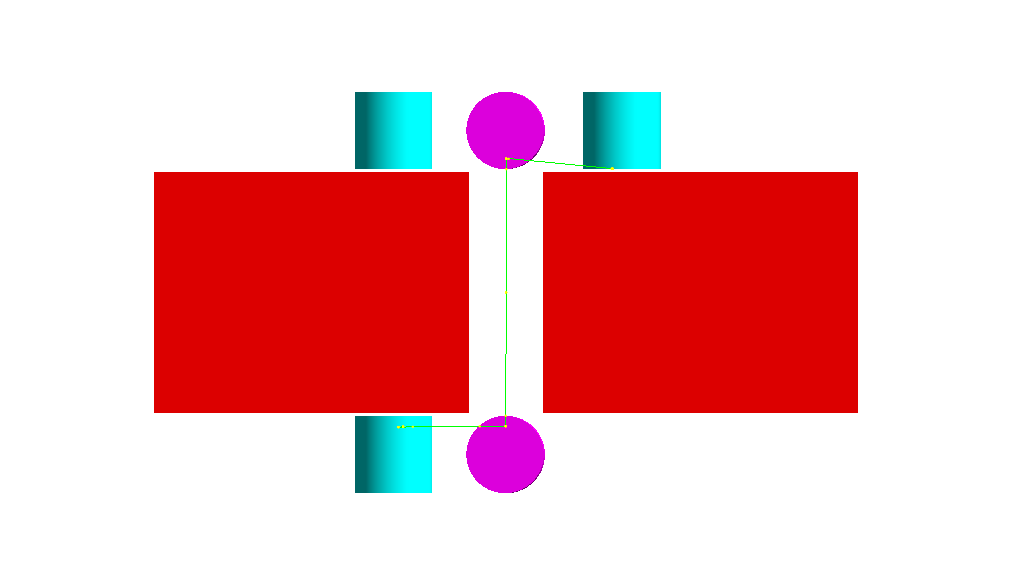}
\caption{Experimental layout in Geant4 simulation. The scatterers are shown in purple, the detectors in teal, and the shielding in violet. The photon paths are shown in green.\label{fig:simulationlayout}}
\end{figure}

Due to the finite solid angles subtended by the detectors and
scatterers to each other and their finite extents, it is possible that
some counting asymmetry may arise purely due to geometry. To better
understand the results of the experiment, a simulation of the
experimental set up has been produced using the Geant4 simulation
toolkit~\cite{Agostinelli:2002hh} and is shown in
Figure~\ref{fig:simulationlayout}. To estimate the contribution of
apparent asymmetry due to geometry effects to the observed overall
asymmetry, the \texttt{G4EmLivermorePhysics} physics list was used,
which neglects the effects of photon polarisation. The apparent
asymmetry due to geometry was estimated to be $R_0 = 1.39 \pm 0.05$,
indicating that the observed asymmetry is enhanced by the geometry. To
reduce the influence of this apparent asymmetry on the observed
asymmetry two variations of the experimental set-up were explored and
are presented in Sections~\ref{sec:Variant1} and~\ref{sec:Variant2}.

To demonstrate the independence of the measured asymmetry from any
interference, the measurement was repeated with the distance between
the scatterers increased by a factor of three. This, correspondingly,
decreases the solid angles subtended by each of the scatterers. The
measurement was carried out for approximately three times longer and a
value of $R = 1.90\pm0.19$ was obtained, while the apparent
geometrical asymmetry was estimated to be $R_0=1.36 \pm 0.07$.

\section{Variations on the Experimental Arrangement}

\subsection{Variation 1}
\label{sec:Variant1}
\begin{figure}[h!]
\centering
\includegraphics[width=0.5\textwidth]{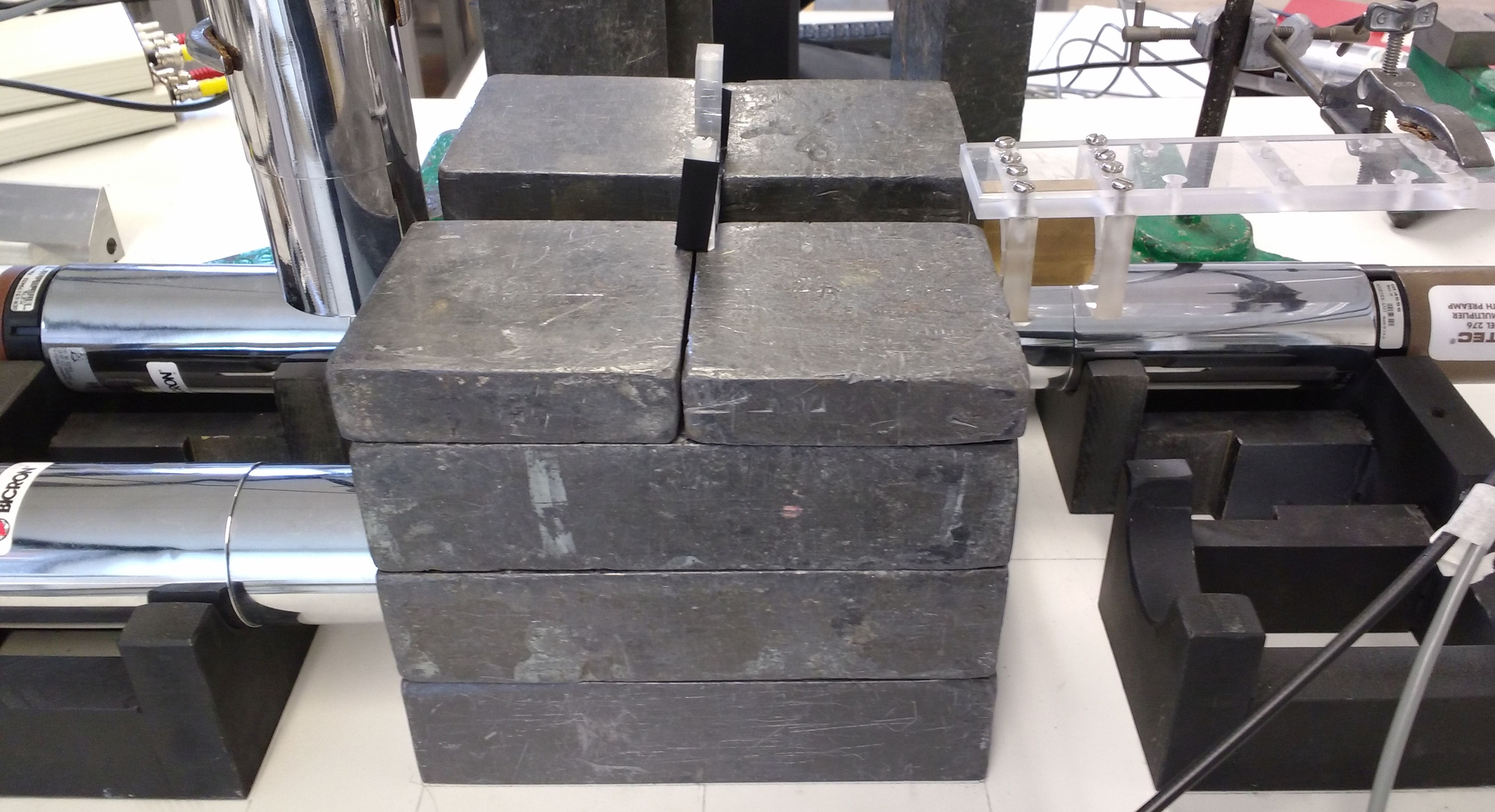}
\caption{Variation 1 of the experiment. The scatterer made of brass is visible on the right-hand side of the photograph.\label{fig:BrassSetup}}
\end{figure}

\begin{figure}[h!]
\centering
\subfigure[\label{fig:Variant1_Results1}]{\includegraphics[width=0.45\textwidth]{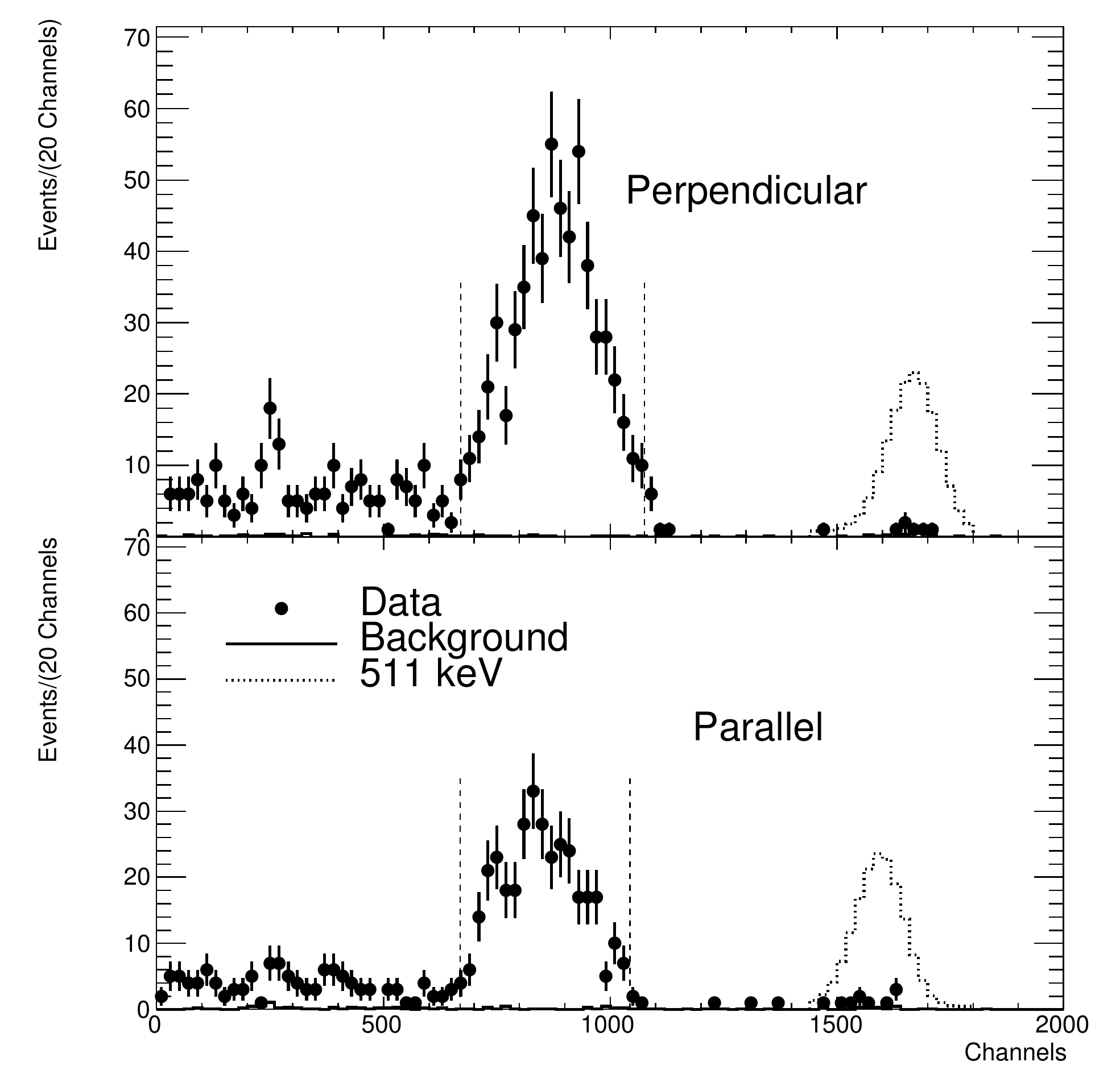}}
\subfigure[\label{fig:Variant1_Results2}]{\includegraphics[width=0.45\textwidth]{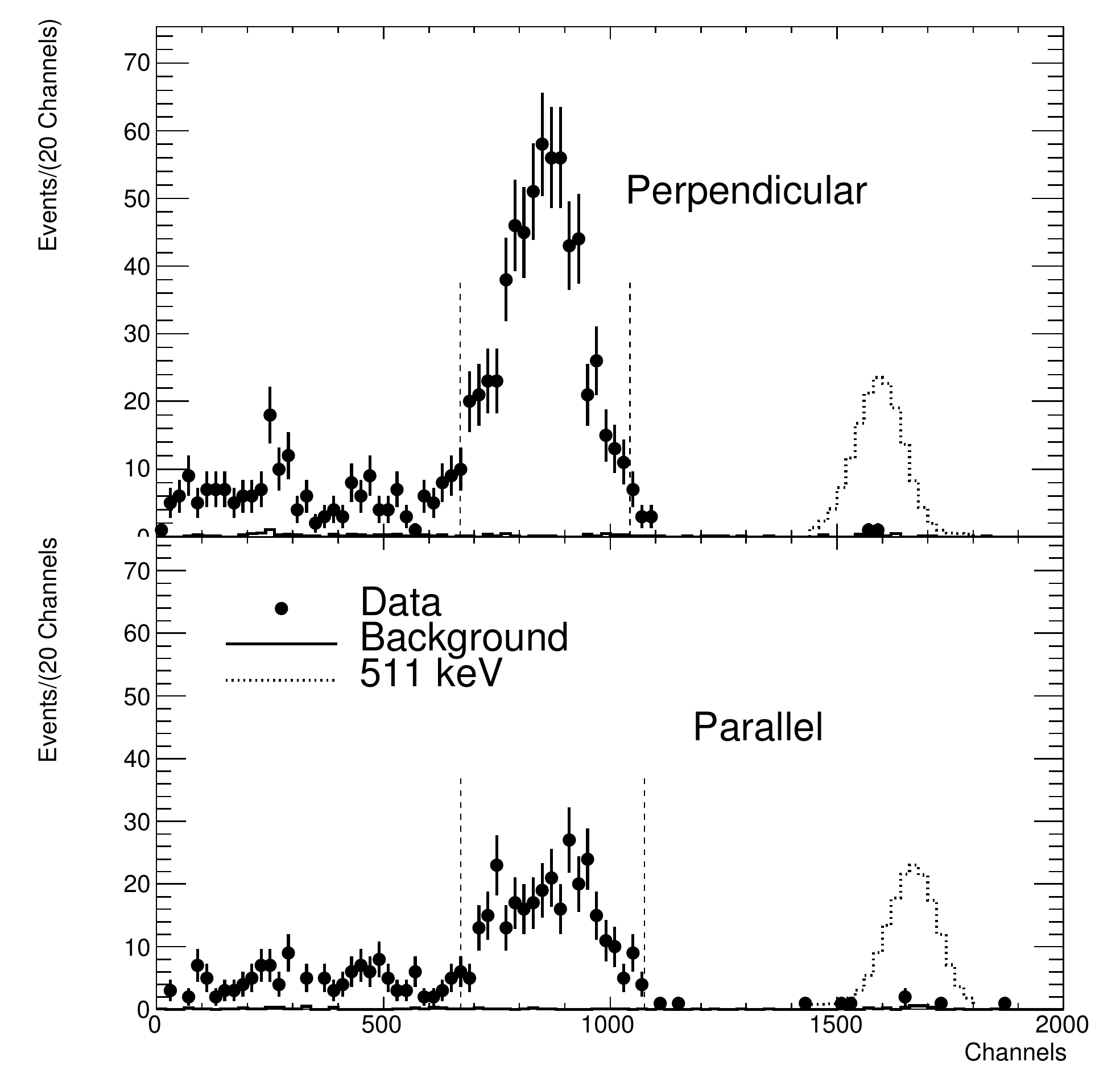}}
\caption{Results showing the asymmetry observed in the orthogonal planes for the first variant setup for \subref{fig:Variant1_Results1} Detector-0 in the initial location and \subref{fig:Variant1_Results2} when the location of Detector-0 was changed.\label{fig:resultsBrass}}
\end{figure}

In this variation of the experiment, shown in Figure~\ref{fig:BrassSetup}, Scatterer-1 is replaced with a rectangle made of brass with dimensions 1.0 cm $\times$ 3.3 cm $\times$ 5.1 cm. From a Geant4 simulation it was determined that for this arrangement the apparent geometric asymmetry was substantially reduced to $R_{0}=1.04\pm0.01$.
The orientation of the brass reduces the range of angles through which a photon can Compton scatter into the detectors. The results of this are shown in Figure~\ref{fig:Variant1_Results1} and give $R_1=1.77\pm0.12$. The measurement was repeated after changing the position of Detector-0 yielding $R_2=2.05\pm0.14$, shown in Figure~\ref{fig:Variant1_Results2}. These data were collected over a period of 22 hours.   

\subsection{Variation 2}
\label{sec:Variant2}
\begin{figure}[h!]
\centering
\subfigure[\label{fig:Variant2_Results1}]{\includegraphics[width=0.40\textwidth]{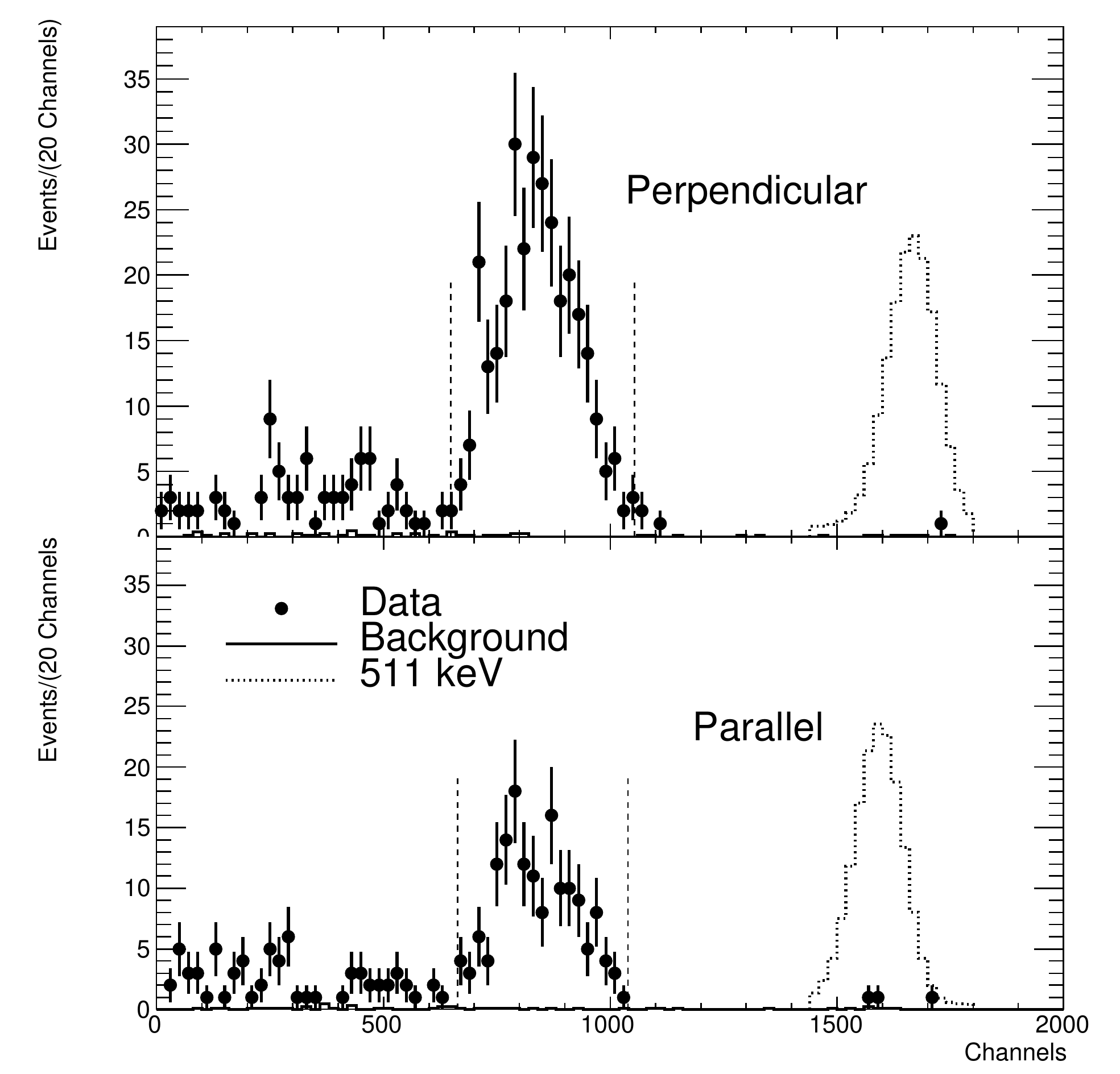}}
\subfigure[\label{fig:Variant2_Results2}]{\includegraphics[width=0.40\textwidth]{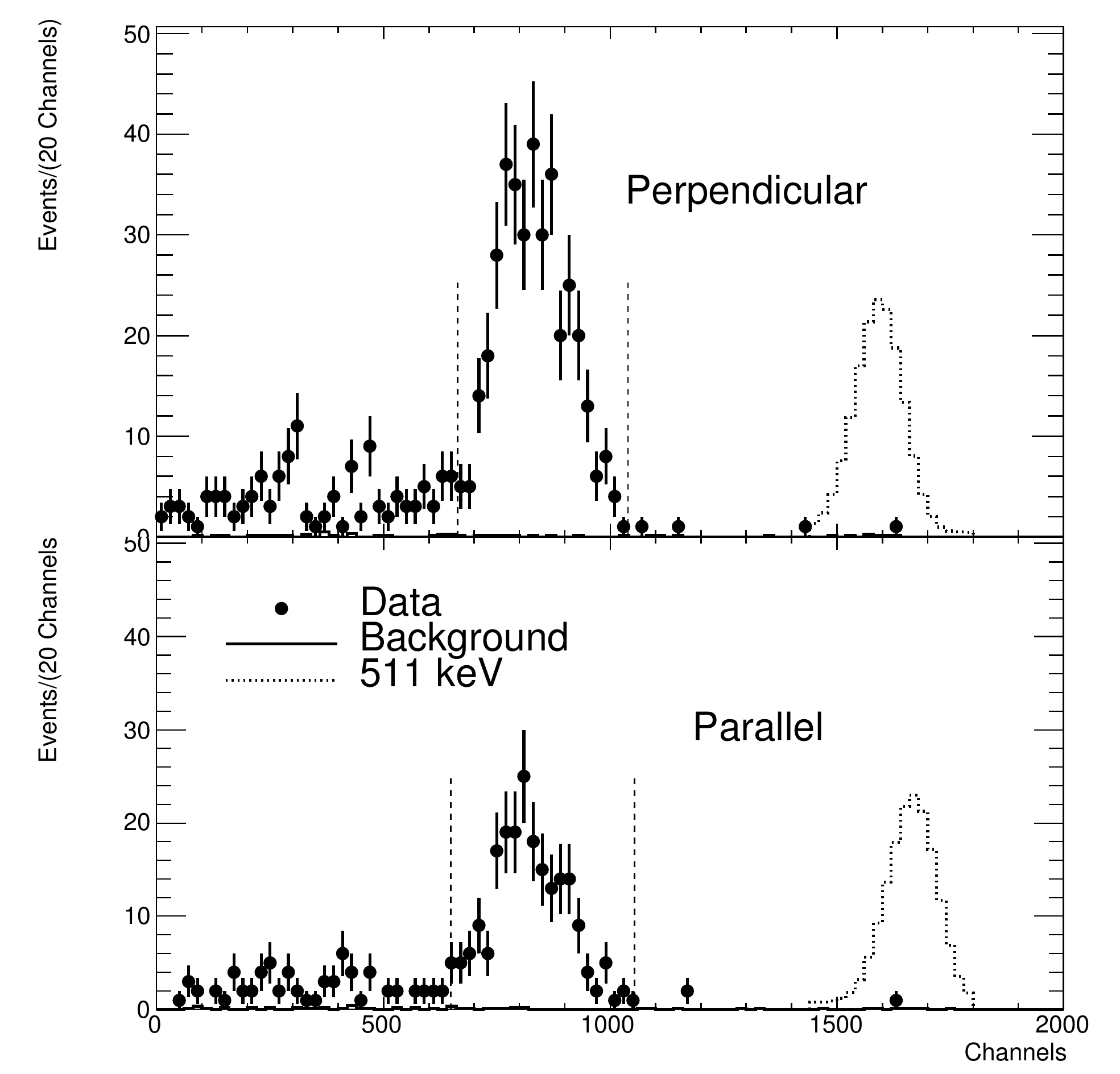}}
\caption{Results showing the asymmetry observed in the orthogonal planes for the second variant setup for \subref{fig:Variant2_Results1} Detector-0 in the initial location and \subref{fig:Variant2_Results2} when the location of Detector-0 was changed.\label{fig:resultsBrassAndScat}}
\end{figure}

Variation 1 is further modified by replacing Scatterer-2 with a 1'' NaI(Tl) scintillator. A Geant4 simulation of this set up gave an apparent asymmetry due to geometry of $R_{0}=1.08\pm0.02$. 
The results of 24 hours of data collection are shown in Figure~\ref{fig:resultsBrassAndScat} for the two repetitions of the experiment. Values of \textbf{$R_1=1.94\pm0.19$} and \textbf{$R_2=1.79\pm0.16$} were obtained.

\section{Discussion}
The presented experiment enables the observation of a correlation in
polarisation between entangled annihilation photons by using the
Compton effect. The experiment was carried out using a positron source
available in the undergraduate laboratory. Data were collected over
the course of a day or overnight. The results in each configuration
were compatible and are shown in Table~\ref{tab:SummaryResults}. 

\begin{table}[h]
\centering
\caption{The experimental results for the initial experiment and the two variants. The averaged asymmetry between the two measurements is given with $\bar{R}$. Also given are the results on the apparent asymmetry, $R_0$, due to geometry obtained with Geant4 simulations for each experimental arrangement. The expected asymmetry $R_{\rm the}$ is also shown, which accounts for the finite size of the detectors in each case. 
\label{tab:SummaryResults}}
\begin{tabular}{l|c|c|c|c|c|c}
Experiment     & $R_1$             & $R_2$              & $\bar{R}$        & $R_{0}$    &$\bar{R}/~R_{0}$ & $R_{\rm the}$  \\ \hline
Initial Set Up & 2.35$\pm$0.19 & 1.99$\pm$0.15  & 2.17$\pm$0.12 & 1.39$\pm$0.05 &1.56$\pm$0.10 & 1.8\\
Variation 1      & 1.77$\pm$0.12 & 2.05$\pm$0.14 & 1.91$\pm$0.09 & 1.04$\pm$0.01 &1.84$\pm$0.09 & 1.9\\
Variation 2      & 1.94$\pm$0.19 & 1.79$\pm$0.16  & 1.87$\pm$0.12 & 1.08$\pm$0.02 & 1.73$\pm$0.12 & 1.9
\end{tabular}
\end{table}

Background was suppressed through coincidence requirements between the
scatterers and detectors.  In the initial experiment both scatterers
were active. However, due to the dimensions of the crystals an
apparent asymmetry arising from the geometry is observed. This is
corrected for in the column labeled $\bar{R}/~R_{0}$ in
Table~\ref{tab:SummaryResults} where the ratio of the observed
asymmetry to the expected geometric asymmetry is taken. Subsequently,
the experiment was modified to reduce this geometric asymmetry. The
results of Variations 1 and 2 of the experiment show that an overall
asymmetry is still readily observable, while asymmetries due to purely
geometric effects are estimated by Geant4 simulations to be
significantly reduced.

Further to the improvements to the experimental arrangement the
detectors subtend finite angles. It is estimated by means of the
Geant4 simulation that for Variant 2 these are $\pm 20^\circ$ in
$\theta$ and $\pm 25^\circ$ in $\phi$, leading to an expected
asymmetry of 1.9. The expected asymmetries in each variant are given
in the right-most column of Table~\ref{tab:SummaryResults}.

Another potential source of asymmetry, unrelated to the photon angular
correlation, is a difference in the efficiencies of Detector-1 and
Detector-2. Although it was experimentally verified that the two
detectors had comparable efficiencies, within the statistical
uncertainty of the asymmetry measurement, data were collected with
Detector-0 moved to the location shown as a dashed cylinder in
Figure~\ref{fig:setupSketch}. This effectively interchanges Detector-1
and Detector-2 as parallel and perpendicular.

Studying the background suppression from coincidence requirements it
was shown that Scatterer-2 could be replaced by a passive scatterer
without an appreciable increase in background. The main experiment
employed four active scintillators. The first variant of the
experiment demonstrates that an asymmetry is still observable when
Scatterer-2 is replaced by an passive scatterer. The choice of size
for Scatterer-2 can also be altered as demonstrated by the second
variant, provided an appropriate amount of time is allocated for data
collection. The experiment could be further simplified to three
scintillators by moving a single detector between the positions of
Detector-2 and Detector-1. This removes the need to ensure that
Detector-1 and Detector-2 have the same efficiency, however, in the
interest of time it was deemed useful to take data in parallel.

\section{Summary}
An experiment to demonstrate correlation of the polarisation of annihilation photons for the undergraduate laboratory is presented. A positron source was used to produced pairs of 511 keV photons travelling at 180$^\circ$. These quantum entangled photons undergo Compton scatterings to produce asymmetries in numbers scattered to orthogonal planes. Lead shielding and coincidence requirements were implemented to reduce background events. Finite geometry effects and apparent asymmetries were discussed and several variations of the experiment were carried out to minimise such effects. 
The experiment provides a demonstration of the Compton scattering dependence on photon polarisation and of quantum entanglement while utilising experimental techniques such as coincidence circuits and background estimation in addition to the standard skills enphasised in the undergraduate laboratory.

\ack This work was performed at the Year 3 Nuclear Physics Laboratory
of the School of Physics and Astronomy at the University of
Birmingham, where this exercise is now offered to undergraduate
students. PK acknowledges support from the School of Physics and
Astronomy during his summer placement. FR acknowledges support from
the Ogden Trust Summer Internship program.

\section*{References}
\bibliographystyle{ieeetr}
\bibliography{annihilation}

\begin{thebibliography}{10}

\bibitem{Einstein1935}
A.~Einstein, B.~Podolsky, and N.~Rosen, ``Can quantum-mechanical description of
  physical reality be considered complete?,'' {\em Phys. Rev.}, vol.~47,
  pp.~777--780, May 1935.

\bibitem{PhysRevLett.28.938}
S.~J. Freedman and J.~F. Clauser, ``Experimental test of local hidden-variable
  theories,'' {\em Phys. Rev. Lett.}, vol.~28, pp.~938--941, Apr 1972.

\bibitem{PhysRevLett.47.460}
A.~Aspect, P.~Grangier, and G.~Roger, ``Experimental tests of realistic local
  theories via bell's theorem,'' {\em Phys. Rev. Lett.}, vol.~47, pp.~460--463,
  Aug 1981.

\bibitem{PhysRevLett.49.1804}
A.~Aspect, J.~Dalibard, and G.~Roger, ``Experimental test of bell's
  inequalities using time-varying analyzers,'' {\em Phys. Rev. Lett.}, vol.~49,
  pp.~1804--1807, Dec 1982.

\bibitem{Wheeler1946}
J.~A. Wheeler, ``Polyelectrons,'' {\em Annals of the New York Academy of
  Sciences}, vol.~48, pp.~219--238, 1946.

\bibitem{Dirac1930}
P.~Dirac, ``On the annihilation of electrons and protons,'' {\em Mathematical
  Proceedings of the Cambridge Philosophical Society}, vol.~26, pp.~361--375,
  1930.

\bibitem{Snyder1948}
H.~S. Snyder, S.~Pasternack, and J.~Hornbostel, ``Angular correlation of
  scattered annihilation radiation,'' {\em Phys. Rev.}, vol.~73, pp.~440--448,
  1948.

\bibitem{Compton:1923zz}
A.~H. Compton, ``{A Quantum Theory of the Scattering of X-rays by Light
  Elements},'' {\em Phys. Rev.}, vol.~21, pp.~483--502, 1923.

\bibitem{Wu:1950zz}
C.~S. Wu and I.~Shaknov, ``{The Angular Correlation of Scattered Annihilation
  Radiation},'' {\em Phys. Rev.}, vol.~77, pp.~136--136, 1950.

\bibitem{Kasday1975}
L.~R. Kasday, J.~Ullman, and C.~S. Wu, ``Angular correlation of
  compton-scattered annihilation photons and hidden variables,'' {\em Il Nuovo
  Cimento B}, vol.~25, pp.~633--661, 1975.

\bibitem{Wilson1976}
A.~R. Wilson, J.~Lowe, and D.~K. Butt, ``Measurement of the relative planes of
  polarization of annihilation quanta as a function of separation distance,''
  {\em Journal of Physics G: Nuclear Physics}, vol.~2, p.~613, 1976.

\bibitem{Pryce1947}
M.~H.~L. Pryce and J.~C. Ward, ``Angular correlation effects with annihilation
  radiation,'' {\em Nature}, vol.~160, pp.~435--435, 1947.

\bibitem{Hetfleis2017}
J.~Hetfleis {\em et~al.}, ``{Entangled $\gamma$-photons - classical laboratory
  exercise with modern detectors},'' {\em Eur. J. Phys}, vol.~39, p.~025403,
  2018.

\bibitem{Dehlinger2002}
D.~Dehlinger and M.~W. Mitchell, ``Entangled photon apparatus for the
  undergraduate laboratory,'' {\em Am. J. Phys.}, vol.~70, p.~898, 2002.

\bibitem{Mitchell2002}
D.~Dehlinger and M.~W. Mitchell, ``Entangled photon apparatus for the
  undergraduate laboratory,'' {\em Am. J. Phys.}, vol.~70, p.~903, 2002.

\bibitem{Bartlett1964a}
A.~A. Bartlett, ``Compton effect: A simple laboratory experiment,'' {\em
  American Journal of Physics}, vol.~32, no.~2, pp.~127--134, 1964.

\bibitem{Bartlett1964d}
A.~A. Bartlett, J.~H. Wilson, O.~W. Lyle, C.~V. Wells, and J.~J. Kraushaar,
  ``Compton effect: an experiment for the advanced laboratory,'' {\em American
  Journal of Physics}, vol.~32, no.~2, pp.~135--142, 1964.

\bibitem{French1965}
W.~R. French, ``Precision compton-effect experiment,'' {\em American Journal of
  Physics}, vol.~33, no.~7, pp.~523--527, 1965.

\bibitem{Stamatelatos1972}
M.~Stamatelatos, ``Compton scattering experiment,'' {\em American Journal of
  Physics}, vol.~40, no.~12, pp.~1871--1872, 1972.

\bibitem{Knights:2017akf}
P.~Knights, F.~Ryburn, G.~Tungate, and K.~Nikolopoulos, ``{Studying the effect
  of Polarisation in Compton scattering in the undergraduate laboratory},''
  {\em Eur. J. Phys.}, vol.~39, no.~2, p.~025203, 2018.

\bibitem{Agostinelli:2002hh}
S.~Agostinelli {\em et~al.}, ``{GEANT4: A Simulation toolkit},'' {\em Nucl.
  Instrum. Meth.}, vol.~A506, pp.~250--303, 2003.

\bibitem{sonzogni2007nndc}
A.~Sonzogni, ``{NNDC} chart of nuclides,'' in {\em International Conference on
  Nuclear Data for Science and Technology}, pp.~105--106, EDP Sciences, 2007.

\end{thebibliography}
\end{document}